\RequirePackage{fix-cm}
\documentclass[aps,prb,amsmath,amssymb,floatfix,reprint,superscriptaddress]{revtex4-2}
\usepackage{graphicx}
\usepackage{dcolumn}
\usepackage{bm}
\usepackage{siunitx}
\usepackage[utf8]{inputenc}
\usepackage[T1]{fontenc}
\usepackage{mathptmx}
\usepackage{color}
\usepackage{nicefrac}
\usepackage{hyperref}
\definecolor{hotblue}{RGB}{12,0,255}
\hypersetup
        {
            colorlinks=true,
            linkcolor=hotblue,
            citecolor=hotblue,
            urlcolor=hotblue,
        }
\begin{document}

\title{Photoneutralization of charges in GaAs quantum dot based entangled photon emitters} 

\author{Jingzhong Yang}
\thanks{Contributed equally to this work}
\author{Tom Fandrich}
\thanks{Contributed equally to this work}
\author{Frederik Benthin}
\thanks{Contributed equally to this work}
\affiliation{Institut f{\"u}r Festk{\"o}rperphysik, Leibniz Universit{\"a}t Hannover, Appelstra{\ss}e~2, 30167~Hannover, Germany}

\author{Robert Keil}
\altaffiliation[Present address: ]{Fraunhofer-Institut f{\"u}r Angewandte Festk{\"o}rperphysik (IAF), Tullastra{\ss}e 72, 79108~Freiburg, Germany}
\affiliation{Institute for Integrative Nanosciences, Leibniz IFW Dresden, Helmholtzstra{\ss}e~20, 01069~Dresden, Germany}

\author{Nand Lal Sharma}
\affiliation{Institute for Integrative Nanosciences, Leibniz IFW Dresden, Helmholtzstra{\ss}e~20, 01069~Dresden, Germany}

\author{Weijie Nie}
\affiliation{Institute for Integrative Nanosciences, Leibniz IFW Dresden, Helmholtzstra{\ss}e~20, 01069~Dresden, Germany}

\author{Caspar Hopfmann}
\affiliation{Institute for Integrative Nanosciences, Leibniz IFW Dresden, Helmholtzstra{\ss}e~20, 01069~Dresden, Germany}

\author{Oliver G. Schmidt}
\affiliation{Institute for Integrative Nanosciences, Leibniz IFW Dresden, Helmholtzstra{\ss}e~20, 01069~Dresden, Germany}
\affiliation{Material Systems for Nanoelectronics, Technische Universit{\"a}t Chemnitz, 09107~Chemnitz, Germany}
\affiliation{Nanophysics, Faculty of Physics and W{\"u}rzburg-Dresden Cluster of Excellence ct.qmat, TU Dresden, 01062 Dresden, Germany}

\author{Michael Zopf}
\email{michael.zopf@fkp.uni-hannover.de}
\affiliation{Institut f{\"u}r Festk{\"o}rperphysik, Leibniz Universit{\"a}t Hannover, Appelstra{\ss}e~2, 30167~Hannover, Germany}
\author{Fei Ding}
\email{fei.ding@fkp.uni-hannover.de}
\affiliation{Institut f{\"u}r Festk{\"o}rperphysik, Leibniz Universit{\"a}t Hannover, Appelstra{\ss}e~2, 30167~Hannover, Germany}
\affiliation{Laboratorium f{\"u}r Nano- und Quantenengineering, Leibniz Universit{\"a}t Hannover, Schneiderberg 39, 30167 Hannover, Germany}

\begin{abstract}
Semiconductor-based emitters of pairwise photonic entanglement are a promising constituent of photonic quantum technologies. They are known for the ability to generate discrete photonic states on-demand with low multiphoton emission, near-unity entanglement fidelity and high single photon indistinguishability. However, quantum dots typically suffer from luminescence blinking, lowering the efficiency of the source and hampering their scalable application in quantum networks. In this paper, we investigate and adjust the intermittence of the neutral exciton emission in a GaAs/AlGaAs quantum dot under two-photon resonant excitation of the neutral biexciton. We investigate the spectral and quantum optical response of the quantum dot emission to an additional wavelength tunable gate laser, revealing blinking caused by the intrinsic Coulomb blockade due to charge capture processes. Our finding demonstrates that the emission quenching can be actively suppressed by controlling the balance of free electrons and holes in the vicinity of the quantum dot and thereby significantly increasing the quantum efficiency by \SI{30}{\percent}.
\end{abstract}

\maketitle 
\section{Introduction}
Generating high-quality single photons and entangled photon states is considered a cornerstone for the development of photonic quantum technologies, e.g, quantum computing \cite{Zhong1460}, quantum communication \cite{Yin_2016}, or quantum simulation \cite{Wang_2019}. Although many mainstream demonstrations of quantum optical applications are implemented with spontaneous parametric down-conversion sources (SPDCs) in the past decades \cite{Chen_2017, Li_2019}, they suffer from a fundamental trade-off between excitation efficiency ($\eta_{ex}$) and multiphoton emission probability, thereby encouraging the search of alternative sources. Semiconductor quantum dots (QDs), referred to as artificial atoms, not only allow for an on-demand generation of high-purity single photons \cite{Somaschi_2016}, they also show high performance in terms of pairwise entanglement fidelity \cite{Huber_2018}, large radiative decay rates and internal quantum efficiencies in contrast to other quantum systems (e.g, atoms and diamond defect centers) \cite{K_rber_2017,Abobeih_2018}. However, an imperfect brightness and luminescence intermittence inhibits the application of such a source in quantum information processing schemes \cite{Deng_2019,you2021quantum,J_ns_2017}.


Tremendous efforts are invested to efficiently extract photons from the high refractive index matrix of QD based devices by modifying the photonic environment, e.g, by broadband microlenses \cite{Schlehahn_2015, Gschrey_2015}, microcavities such as circular Bragg gratings \cite{Liu_2019} and micropillars \cite{Wang_2019_Ondemand}, or metal antennas \cite{Bigourdan_2014, Galal_2017} based on directional emission enhancement from surface plasmon resonance. The overall brightness of QD based photon sources is further influenced by the excitation efficiency, which corresponds to the probability of QDs being occupied at the desired state under resonant excitation \cite{Santori_2004, Huber_2016}.
The degradation of excitation efficiency is typical for solid-state quantum emitters like single colloidal nanocrystals \cite{Nirmal_1996,Efros_2016} or organic molecules \cite{Hoogenboom_2007}, manifesting as strong intermittency phenomena in the photon emission (called "blinking"). Even for \uppercase\expandafter{\romannumeral3}-\uppercase\expandafter{\romannumeral5} QDs, epitaxially grown in an ultrahigh vacuum atmosphere, it is impossible to fully avoid this effect without embedding QDs in devices such as n-i-p diode structures, which would allow for charge tuning \cite{Somaschi_2016,Zhai_2020}. An all-optical approach to stabilize the solid-state charge environment for a resonantly excited QD is the concept of photoneutralization of charges by illumination with an additional weak gate laser, which has been investigated in self-assembled InAs/GaAs QDs \cite{Nguyen_2012,Nguyen_2013,Huber_2016}. In recent years, the emerging family of GaAs/AlGaAs QDs \cite{Keil_2017,Huber_2017} has shown intriguing features that benefit a high-quality entangled photon emission close to the rubidium D2 transitions~\cite{Akopian_2011,Jahn_2015}. These include a high QD symmetry and homogeneity, low electron-nuclear spin hyperfine interaction~\cite{Welander2015}, and reduced light- and heavy hole mixing~\cite{Huo_2013}.


Here we  investigate the effect of a weak gate laser on entangled photon sources based on GaAs/AlGaAs quantum dots. For the generation of entangled photon pairs and investigating photoneutralization effects, we employ resonant two-photon pumping (TP) excitation to coherently drive the biexciton state in GaAs/AlGaAs quantum dots. Strong blinking and an additional emission from the positive exciton ($X^+$) transition are observed, indicating a degraded TP excitation efficiency. This effect is counteracted with a second, wavelength tunable gate laser in the low-power regime, allowing control over the charge carrier exchange rates between the barrier material and the QD. The impurity induced excess of holes in the QD ground state is therefore suppressed. Using photoluminescence excitation (PLE) spectroscopy and second-order correlation measurements, we probe the continuum states in the barrier, and control the corresponding charging rate of the QD. 
\section{Methods}
The GaAs/AlGaAs QDs under investigation are grown by molecular beam epitaxy with the in-suit droplet etching and nanohole infilling method \cite{Keil_2017,da2021gaas}. After the Al-droplet etching process on the \SI{235}{\nano\meter} thick $Al_xGa_{1-x}As$ (x=0.15), a layer of \SI{2}{\nano\meter} GaAs is deposited to fill the nanoholes, followed by \SI{200}{\nano\meter} of $Al_xGa_{1-x}As$.  In order to efficiently extract the photons out of the host material, a dielectric antenna device is fabricated \cite{Chen_2018}. Excitation of QDs is accomplished by illumination with \SI{80}{\MHz} Ti-Sapphire pulsed laser light, with a pulse length of $\sim \SI{9}{\pico\second}$. To realize TP excitation, the energy of the laser pulse is adjusted exactly at the half energy of the biexciton state $\left | XX \right \rangle $. Additionally, a weak continuous wave (cw) laser (FWHM$\approx \SI{100}{\kHz}$) is applied with power around hundreds of \SI{}{\nano\watt} for photoneutralization.

\begin{figure}[htp]
    \centering
    \includegraphics[width=0.45\textwidth]{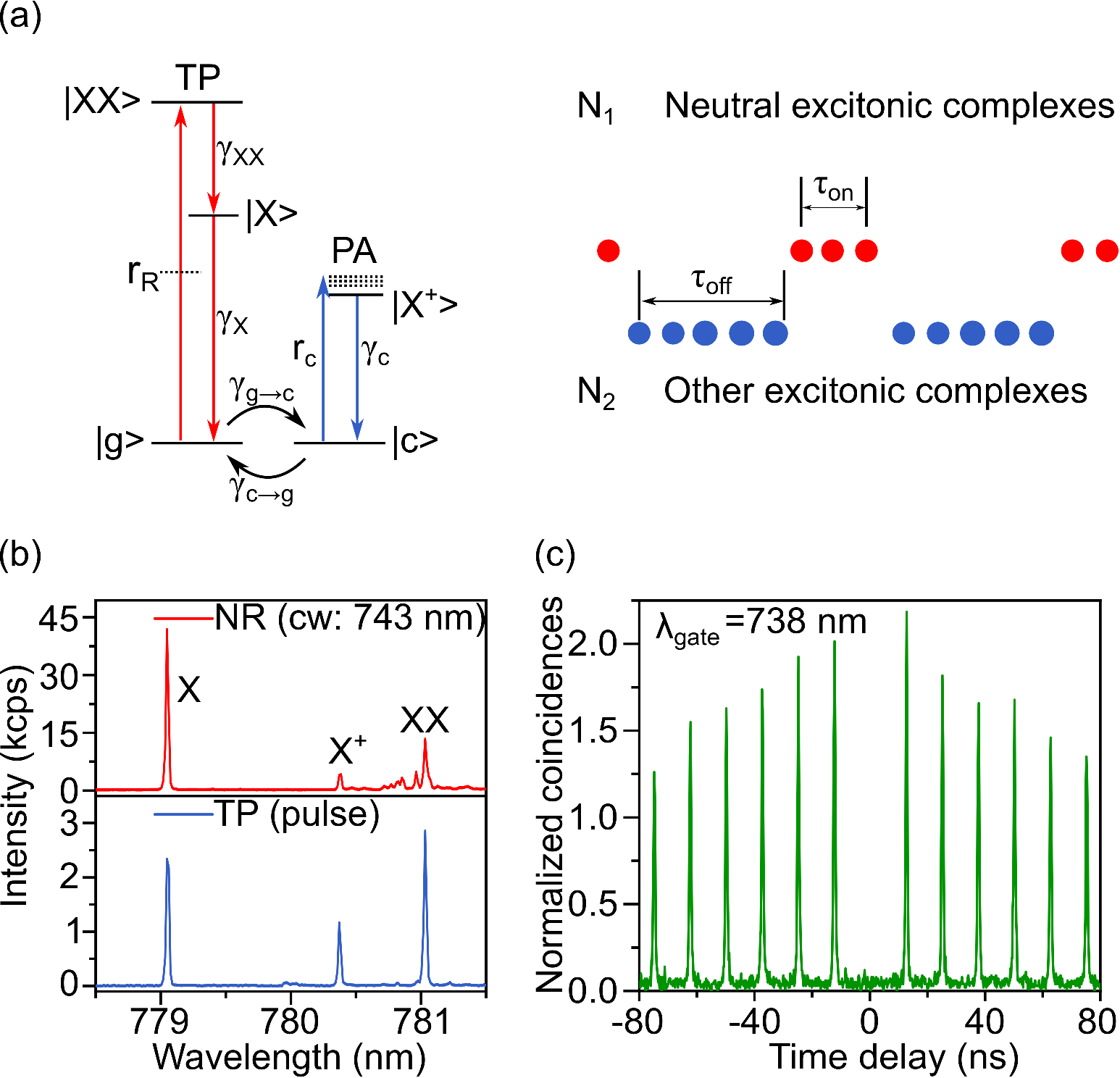}
    \caption{Characteristic blinking behavior of the QD emission under optically gated TP excitation. (a) Schematic of the charge carrier transition dynamics (left) and the effect on the emitted single photon stream (right). (b) Photoluminescence spectra of an exemplary QD under NR and TP excitation with cw and pulsed laser, respectively, showing the neutral exciton (X) and biexciton (XX) as well as the positively charged trion (X$^+$) emission. (c) Normalized autocorrelation histogram of $X$ photons under TP excitation with an additional gate laser at \SI{738}{\nano\meter}.}
    \label{fig:overview}
\end{figure}

The diagram on the left in Fig. \ref{fig:overview}(a) illustrates the QD as a two and three-level system including the dynamics of charge capture processes related to the blinking behavior under optically gated TP excitation. If the QD is initially empty, described by the ground state $\left | g \right \rangle $, it can be pumped to the biexciton state $\left | XX \right \rangle $ using a TP excitation scheme with a pumping rate $r_R$ (dashed line indicates the virtual energy level in TP excitation). Then, $\left | XX \right \rangle $ starts to decay via the exciton state $\left | X \right \rangle $ to the ground state by emitting a pair of polarization-entangled photons with spontaneous decay rates of $\gamma_{XX}$ and $\gamma_X$, respectively. However, residual electrons and trapped by impurities (i.e. carbon acceptors) in the vicinity of the QDs can tunnel into the QDs with rates of $\gamma_e^{(d)}$ and $\gamma_h^{(d)}$ (not shown in the diagram) resulting in the charged ground state $\left | c \right \rangle $. The formation of the neutral biexciton will be blocked by the residual hole in the QD, similar to the Coulomb blockade effect \cite{Nguyen_2013}. According to the model of photoneutralization, holes and electrons are created in the barrier by an additional, weak and off-resonant gate laser with rates of $\gamma_h^{(b)}$ and $\gamma_e^{(b)}$, respectively. It modifies the rates $\gamma_{g\to c}$ and $\gamma_{c\to g}$ describing the transition between the neutral and charged ground state of the QD. The transfer rate between impurities and the QD is negligible if $\gamma_e^{(d)}, \gamma_h^{(d)}\ll \gamma_e^{(b)}, \gamma_h^{(b)}, \gamma_{X}, \gamma_{XX}$. Therefore, the neutral ground state $\left | g \right \rangle $ converts into the charged ground state $\left | c \right \rangle $ with an average number of electrons or holes in steady regime within a time range $\left( \tau_{on}=\nicefrac{1}{\gamma_{g\to c}} \right )$. This leads to the phonon-assisted (PA) excitation of the trion state (i.e. $\left | X^+ \right \rangle $) with a pumping and radiative rate of $r_c$ and $\gamma_c$ \cite{Reindl_2017}. After a while $\left(\tau_{off}=\nicefrac{1}{\gamma_{c\to g}}\right)$, the charged state is neutralized, leaving the QD empty and resulting in TP excitation of the $\left | XX \right \rangle $ state. The right diagram in Fig. \ref{fig:overview}(a) presents the blinking dynamics of the photon emission from the QD within the charge and neutralization process of the ground state. In a time range of $\tau_{on}$, the QD stays neutral so that single photons pairs are continuously emitted from the spontaneous transition processes. But once the ground state is charged, neutral exciton photon emission is quenched and replaced by other excitonic state emission until it is neutralized again.
\section{Results and Discussion}
Fig. \ref{fig:overview}(b) shows the typical non-resonant (NR) and TP excitation photoluminescence (PL) spectra of a single QD, in which three dominant transitions of excitonic states are distinguished. By comparing the two spectra, it can be seen that the $\left | X^+ \right \rangle$ state exists in TP excitation while the further red-shifted, weak transition lines around $\left|XX\right \rangle$ disappear. It demonstrates that the neutral and charged ground states are alternating, giving rise to bright $\left|X^+\right\rangle$ state emission and causing the intermittency of the neutral $X$ photon emission. Fig. \ref{fig:overview}(c) displays the autocorrelation measurement of the neutral $X$ photon emission from the QD under TP excitation using a Hanbury-Brown and Twiss setup. The anti-bunching at zero time delay reveals a pure single photon emission. At the same time, a superimposed bunching at a larger time delay demonstrates the switching of the QD between a neutral and a charged ground state.

\begin{figure*}[ht]
    \centering
    \includegraphics[width=\textwidth]{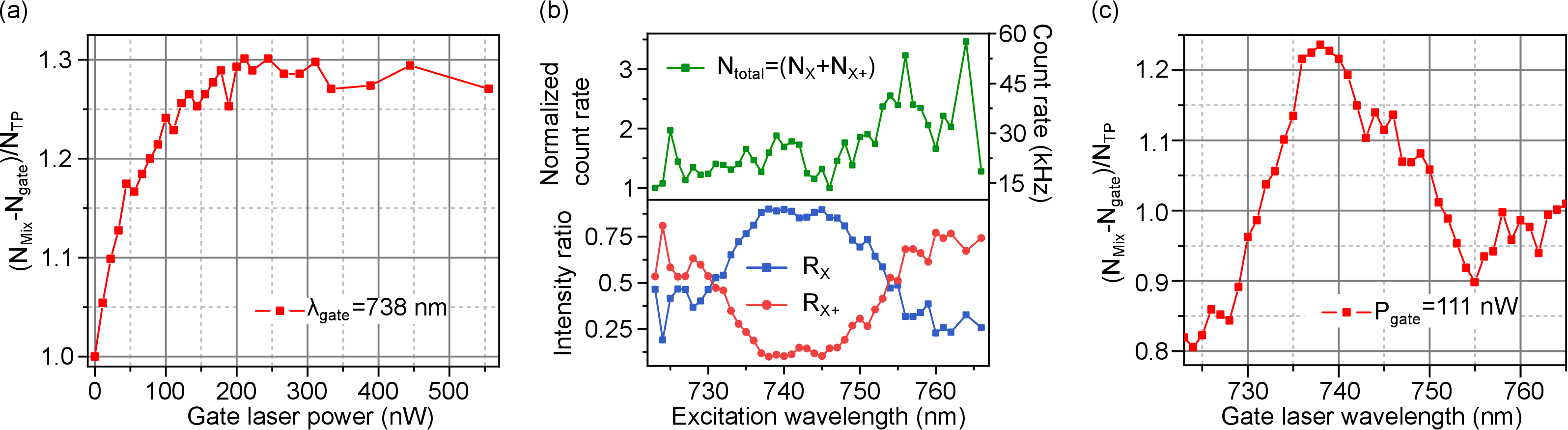}
    \caption{Dependence of QD emission intensity on gate laser power and wavelength. (a) Normalized gate laser power dependent fluorescence intensity of $X$ photons from the QD under TP excitation. (b) Excitation spectroscopy measurement in the absence of TP excitation under  gate laser power $P_{gate}=\SI{6.3}{\micro\watt}$, showing the total intensity and the intensity ratio of the $X$ and $X^+$ transitions ($R_X=N_X/N_{total}$ and  $R_{X^+}=N_{X^+}/N_{total}$). (c) Normalized gate laser wavelength dependent fluorescence intensity of $X$ photons from the QD under TP excitation.}
    \label{fig:intensity}
\end{figure*}

For investigating the influence of the optical gate effect on the blinking, we first study the photoluminescence intensity dependence on the gate laser power and wavelength in the steady state regime. Fig. \ref{fig:intensity}(a) shows the dependence of the neutral $X$ photon counts on the power of the gate laser at a constant power of the pulsed TP excitation laser. To illustrate that the improvement of the intensity arises from the neutralization effect induced by the gate laser instead of the simultaneous off-resonant fluorescence, the photon counts emitted from the QD when illuminated by two lasers ($N_{Mix}$) are subtracted by the counts present when solely excited by the gate laser ($N_{gate}$). The difference ($N_{Mix}-N_{gate}$) is then normalized to the $X$ count rate ($N_{TP}$) when only the pulsed TP excitation is present.  The curve increases and eventually saturates at around 1.3 with increasing gate laser power, proving the gate laser effect, aligned with a previous report \cite{Nguyen_2013}. Next, we implement an excitation spectroscopy measurement with the cw laser (Fig. \ref{fig:intensity}(b)) from \SI{723} to \SI{766}{\nano\meter} after the subtraction of the background, to explore how the excitation wavelength affects the excitation efficiency of the different excitonic complexes in the QD. In the upper panel of Fig. \ref{fig:intensity}(b), the total intensity of the $X$ and $X^+$ photons is shown, which keeps constant at wavelengths below \SI{750}{\nano\meter} and then gradually increases. We attribute the increase at $\lambda > \SI{750}{\nano\meter}$ to a higher probability of the QD ground state to be either empty or populated with one residual hole. The probability of a negatively charged ground state or the formation of dark excitonic states is therefore expected to decrease.
The lower panel of Fig. \ref{fig:intensity}(b) shows the intensity ratio of the $\left | X\right\rangle$ and $\left | X^+\right\rangle$ states as a function of the non-resonant laser wavelength. A strong contrast of the intensity ratio is observed at $\lambda \approx \SI{738}{}$ and $\SI{745}{\nano\meter}$. Since the spontaneous decay rates of $X$ and $X^+$ photons are assumed to be comparable \cite{Sch_ll_2019}, we can conclude that the QD tends to be empty at the ground state ($n_h^{(st)}=0$) for these specific wavelengths. To demonstrate that the laser wavelength is able to control the charge state of the QD, we study the gate laser wavelength dependent emission intensity of neutral $X$ photons (Fig. \ref{fig:intensity}(c)) under TP excitation, in which most of the bright states are strictly from $\left|X^+\right\rangle$ and the cascade emission of the $\left|XX\right\rangle$ states. The power of the gate laser is kept constant ($P_{gate}=\SI{111}{\nano\watt}$) below the power at which the optical gate effect saturates ($P_{sat}\sim \SI{200}{\nano\watt}$). We observe a similar intensity fluctuation ($R_x$) as shown in Fig. \ref{fig:intensity}(b), revealing that the QD can be actively photoneutralized.

\begin{figure*}[ht]
    \centering
    \includegraphics[width=\textwidth]{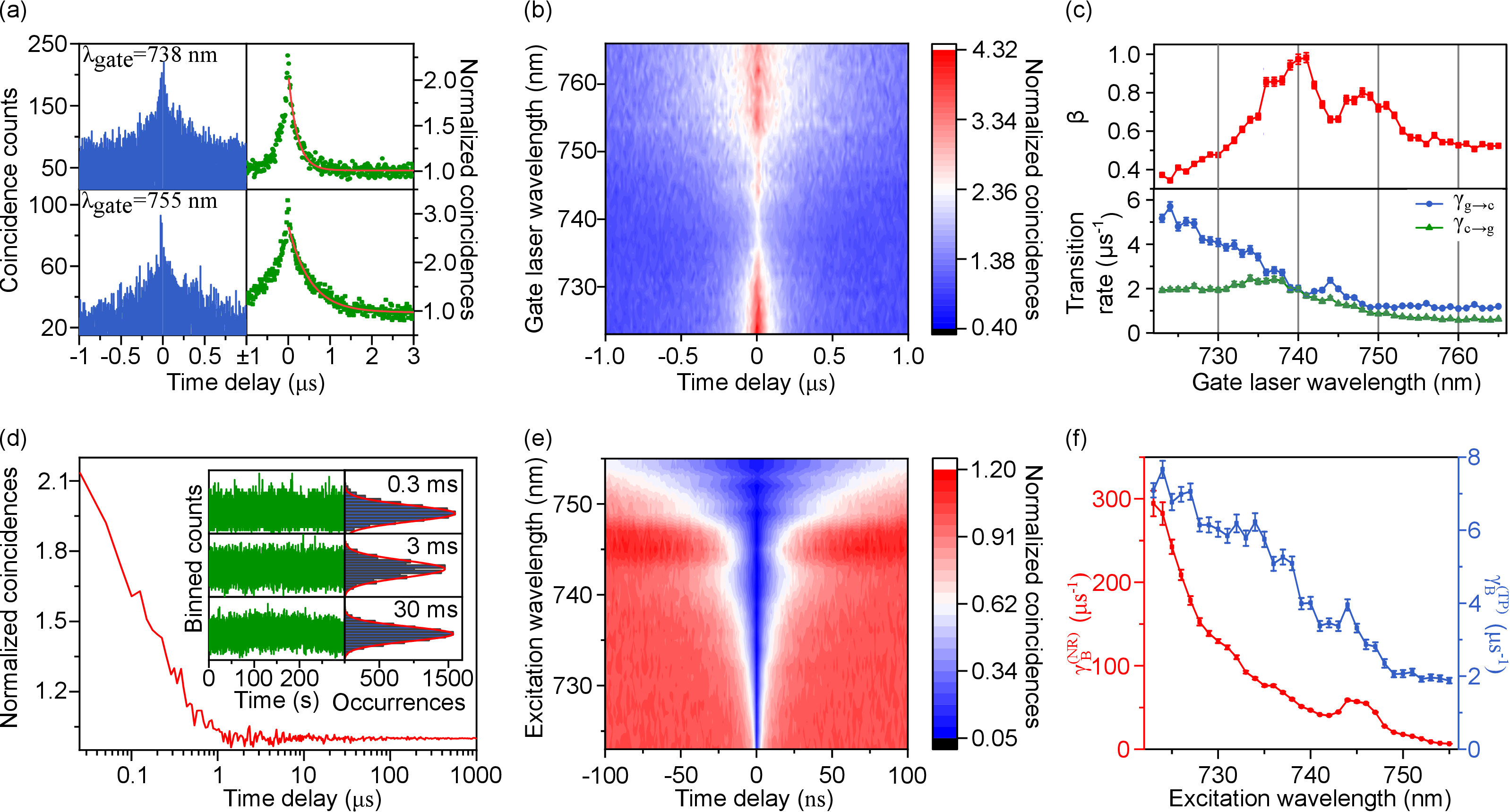}
    \caption{Characterization of the gate laser dependent blinking characteristics of the neutral $X$ and $X^+$ photon emissions. (a) Intensity autocorrelation histograms (left) for $X$ photons revealing pure single-photon emission and a superimposed bunching due to blinking. Normalized histogram (right) with bin size set to the inverse of the laser repetition rate is used to apply an exponential decay model (red line). Measurements at two exemplary gate laser wavelengths are shown. (b) Normalized coincidences as a function of time delay and gate laser wavelength, with fixed powers of the resonant and gate lasers ($P_{gate}=\SI{166}{\nano\watt}$). (c) On-off time ratio $\beta$ (top) and the charge capture rates (bottom) as a function of gate laser wavelength. (d) Normalized intensity autocorrelation of $X$ photons in a logarithmic time delay scale up to one millisecond ($\lambda_{gate}=\SI{738}{\nano\meter}$). (Inset: Time trace (green) and histogram (blue) of detector count rates for \SI{0.3}{\milli\second}, \SI{3}{\milli\second} and \SI{30}{\milli\second} time bins, following Gaussian distribution with the statistics (red). (e) Normalized coincidences from a cross-correlation measurement between $X$ and $X^+$ photons as a function of time delay and excitation laser wavelength. (f) Average blinking rates $\gamma_B$ under TP and NR excitation as a function of the wavelength of the non-resonant excitation laser.}
    \label{fig:correlation}
\end{figure*}

In the following we present the corresponding behavior as a function of gate laser wavelength and explain the blinking phenomena according to the proposed model, as intensity autocorrelation measurements reveal the photon emission properties in the dynamic regime. These correlations are shown in Fig. \ref{fig:correlation}(a) for the $X$ photon emission. It displays two exemplary histograms of the autocorrelation measurements (left) and the corresponding coincidence counts binned according to each laser pulse emission (right), normalized to the coincidences at large time delays. The missing peak in the center of the histograms resulting from anti-bunching is neglected during the binning of coincidence counts in the right graph. The remarkable bunching behavior (green curve) reflects the blinking dynamics for which photons emitted close to each other within the time range ($\left | \tau_B \right |\leq \SI{1}{\micro\second}$) have a higher chance to result in coincidence counts. By comparing the theoretical modeling of these two exemplary bunching curves, one can notice that the blinking dynamics is tuned when changing the gate laser wavelength. It deserves to be mentioned that the slight mismatch of the model with the central data points close to a time delay of $\tau=\SI{0}{\micro\second}$ is 
presumed to be the dark exciton transition \cite{Davan_o_2014}.  Fig. \ref{fig:correlation}(b) shows the gate laser wavelength and time-delay dependent normalized coincidences [each horizontal cut of the data corresponds to a green curve of the binned data in Fig. \ref{fig:correlation}(a)] at different wavelengths in the form of a color-map. The bunching amplitude undergoes a descent and ascent with increasing gate laser wavelength,  reaching the lowest value at $\sim\SI{740}{\nano\meter}$. It implies a decreased blinking, qualitatively validating the behavior observed in Fig. \ref{fig:intensity}(c) about the intensity change as a function of the gate laser wavelength. 

To quantitatively understand the blinking phenomenon related to the capture of charges in the QD, we extract the values of the parameters based on the modeling of the $g^{(2)}$, described by \cite{Jahn_2015,Sallen_2010}, 

\begin{align}
    \label{equ:auto}
    g^{(2)}(\tau,\lambda)&=\left[1+\left(\frac{1}{\beta(\lambda)}\right) e^{-\gamma_B(\lambda)\left | \tau \right |}\right]
\end{align}
\begin{align}
    \label{equ:blink}
    \gamma_B(\lambda)&=\gamma_{g\to c}(\lambda)+\gamma_{c\to g}(\lambda)
\end{align}

with $\beta$ denoting the blinking on-off ratio ($\tau_{on}/\tau_{off}$), $\gamma_B$ the blinking rate ($\gamma_B={\textstyle \sum_{i}^{}\gamma _i}, i=g\to c, c\to g$). The upper panel in Fig. \ref{fig:correlation}(c) shows the average ratio of on and off times as a function of the gate laser wavelength, in which $\beta$ follows exactly the tendency of intensity fluctuations in the steady state (Fig. \ref{fig:intensity} (c)). It verifies our previous assumption that the gate laser wavelength can tune the number of remaining charges in the QDs, thus affecting the excitation efficiency ($\eta_{ex}^{(min)}=\SI{25.57(8)}{}\pm\SI{0.403}{\percent}, \eta_{ex}^{(max)}=\SI{49.53(9)}{}\pm\SI{0.645}{\percent}$). A closer look at the effect of the gate laser on the charging and neutralization process is obtained by deducing the capture rates of holes and electrons as displayed in the lower panel. The charging rate (blue curve) generally stays higher than the neutralization rate (green curve) in the wavelength range from $\SI{723}{}\sim\SI{765}{\nano\meter}$, therefore, the ground state of the QD is charged with an excess of holes for most of the wavelengths. When the gate laser is tuned to $\sim\SI{740}{\nano\meter}$, the two lines meet each other, corresponding to a balanced capture rate of holes and electrons between the barrier and the QD. ($\gamma_{e}^{(b)}=\gamma_{h}^{(b)}$). Simultaneously, the low transition of charges tunnelled from defects to the QD with rates of $\gamma_e^{(d)}, \gamma_h^{(d)}$ dominates. To completely eliminate the excess of charges in the QDs, a slightly higher ejection rate of electrons is preferred, in order to neutralize the holes additionally induced by the impurities.

In Fig. \ref{fig:correlation}(d), the normalized intensity autocorrelation of the $X$ photons is plotted in a logarithmic time scale up to $\SI{1}{\micro\second} \le\left|\tau\right|\le \SI{1000}{\micro\second}$. The first order linear decay at short delays ($\left|\tau\right|\le \SI{1}{\micro\second}$) indicates that there are no other potential dark states causing the blinking of the photon emission \cite{Davan_o_2014}. The curve then remains flat up to $\SI{1}{\milli\second}$, indicating that there is no blinking observed at this time scale. To study the blinking on an even larger time scale, Fig. \ref{fig:correlation}(d) shows the fluorescence intensity over time, under different time bin sizes of \SI{0.3}{}, \SI{3}{} and \SI{30}{\milli\second} (left). From the histogram of the intensity values of each curve (right) it is clear that the intensity distribution exhibits a single Gaussian model behavior, verifying the abscence of blinking at a longer time range $\left|\tau\right|\ge \SI{1}{\milli\second}$.

The total capture rate of charges ($\gamma_{B}$) between QDs and barrier material decreases with increasing gate laser wavelength, and not only under TP excitation. For NR excitation, the charging rate is affected by the excitation laser wavelength, explaining the excitation spectroscopy in Fig. \ref{fig:intensity}(b). The color map in Fig. \ref{fig:correlation}(e) shows the normalized coincidences from a cross-correlation measurement between $X$ and $X^+$ photons as a function of wavelength of the cw excitation laser. The excitation wavelength is limited to values below $\SI{755}{\nano\meter}$, due to the observed laser background at higher wavelengths. Similar to the bunching envelope in the autocorrelation measurement, the here generally symmetric anti-bunching reveals the effect of the laser wavelength on the charge capture rate of the QD from its surrounding environment. The anti-bunching in the cross-correlation can be modelled with the below formula \cite{Sallen_2010,Hopfmann_2021},

\begin{align}
        \label{equ:cross_blink}
        g_{X, X^+}^{(2)}(\tau) &= (1-e^{\gamma_B(\lambda)\left|\tau \right|}) g_{cross}^{(2)}(\tau)
\end{align}
\begin{align}
        \label{equ:cross_anti}
        g_{cross}^{(2)}(\tau) &= [H(\tau) g_{X^+}^{(2)}(\tau)+H(-\tau) g_{X}^{(2)}(\tau)]
\end{align}

where the first term in Eq. (\ref{equ:cross_blink}) is the total capture rate describing a charging of the ground state of the QD with one hole (related $X^+$ photon emission), followed by a return to the neutralized state ($\left| g\right\rangle\to \left| c\right\rangle\to \left| g\right\rangle$). The second term is the conventional anti-bunching, a result from the single photon character of the $X$ and $X^+$ emissions. This is expressed in Eq. (\ref{equ:cross_anti}), where $H$ is the Heaviside step function while $g^{(2)}_{X^+}$ and $g^{(2)}_{X}$ are the anti-bunching functions of $X^+$ and $X$, respectively. At lower excitation wavelengths it can be seen that the anti-bunching shows an asymmetric dip in the center. 
This can be explained by the different rates contributing to the dynamics of $X$ and $X^+$ emission, taking into account the pumping and decay processes under NR excitation. The decay processes (described by $r_{c}, \gamma_{R}, r_{X}, \gamma_{X^+}$) are much faster than the blinking rate of the ground state ($\gamma_B$) \cite{Hopfmann_2021}. With increased wavelength, the asymmetric anti-bunching can not be distinguished anymore because of the low signal-to-noise ratio and the dominating blinking rate, leading to a broader and more symmetric anti-bunching.


Fig. \ref{fig:correlation}(f) shows the comparison of the extracted $\gamma_B$ from the autocorrelation of Fig. \ref{fig:correlation}(b) and the cross-correlation of Fig. \ref{fig:correlation}(e), respectively. Overall, the capture rate of the empty ground state and charged state with one hole is reduced with increasing gate/excitation laser wavelength. However, $\gamma_B$ under NR excitation is much higher than the rate under TP excitation due to the different dynamics and possible formation of other excitonic states (eg. $X^-$ state) in NR excitation.

\begin{figure}[htbp] 
    \centering
    \includegraphics[width=0.45\textwidth]{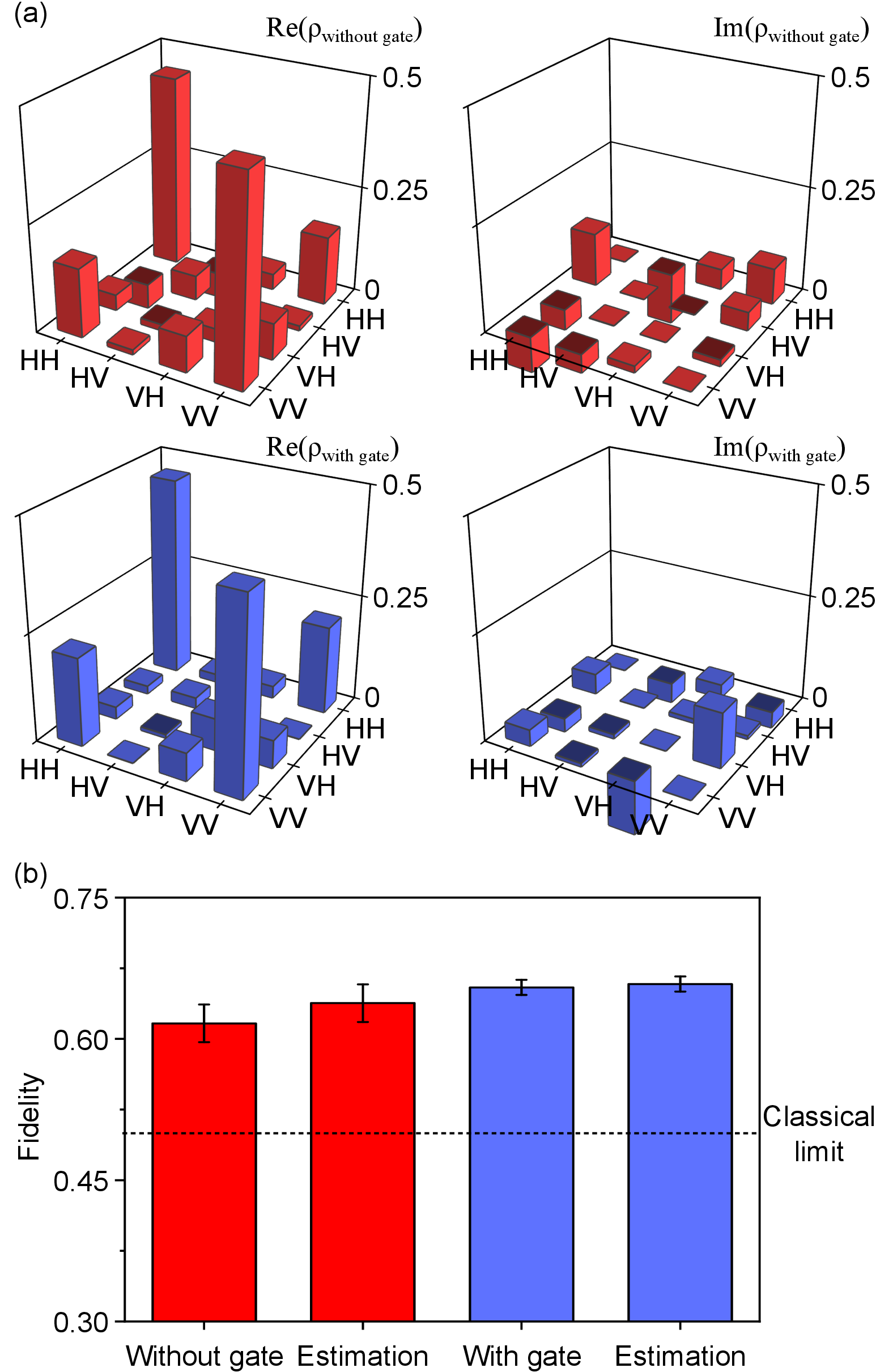}
    \caption{(a) Real (left) and imaginary (right) of reconstructed two-photon density matrices between $XX$ and $X$ photons under TP excitation without (top) and with (bottom) gate laser. (b) Fidelity to the expected Bell state $\left|\Phi^+\right\rangle$ for TP excitation without (red) and with (blue) gate laser.}
    \label{fig:tomography}
\end{figure}

After the investigation of photon-neutralization on the charge environment in the vicinity of QDs, we further try to learn more about the influence of the gate effect on the quality of the polarization-entanglement of photons, by performing the quantum state tomography. Fig. \ref{fig:tomography}(a) presents the polarization-resolved $XX-X$ density matrices measurement for the emission from the same QD under TP excitation without and with gate laser ($\lambda_{gate}=\SI{738}{\nano\meter}, P_{gate}=\SI{111}{\nano\watt}$), respectively. Without any post-selection method, we obtain the raw fidelities of $f_{\text{without gate}}=0.61(6)\pm0.020, f_{\text{with gate}}=0.65(5)\pm0.008$ by projecting the state to the Bell state $\left|\Phi^+\right\rangle$ in Fig. \ref{fig:tomography}(b). Utilizing the maximum-likelihood estimation method, fidelities of $f_{\text{without gate}}=0.63(8)\pm0.020, f_{\text{with gate}}=0.65(8)\pm0.008$ are hereby extracted from the reconstructed density matrices, which are nearly constant. Therefore, we can conclude that the fidelity to the Bell state is not dependent on the excitation scheme \cite{Reindl_2017}.
\section{Conclusions}
To summarize, we have shown that the electronic environment and material impurities play a pivotal role in obtaining an efficient optical response from a semiconductor QD based photon source. Charge capture processes are identified to be the major cause of luminescence intermittence in GaAs/AlGaAs QDs subject to two-photon excitation to the biexciton. The power and wavelength of a gate laser acts on the capture rate of holes and electrons, particularly close to the band gap of the barrier material ($\sim\SI{740}{\nano\meter}$ for $Al_{0.15}Ga_{0.85}As$). An enhancement of up to \SI{30}{\percent} in excitation efficiency is observed, and the entanglement fidelity of the emitted photon pairs is maintained by applying the optical gate. The intensity of the QD emission in the steady state is well explained by investigating the charge capture dynamics with auto-/cross- correlation measurements from NR and TP excitation schemes. Our finding demonstrates that photo-generated charge carriers modify the electronic environment of the QDs and thus increase the efficiency of the source while maintaining the degree of photonic entanglement. One possibility to gain further control of the charge in the QD is embedding it in n-i-p type charge tunable devices\cite{Zhai_2020}. Since such devices also increase the fabrication complexity, especially in combination with nanostructure fabrication \cite{Liu_2019,Wang_2019_Ondemand} or, e.g., strain-field tuning \cite{Chen_2016, Trotta_2016}, an all optical and easily implementable method to control the charge population of the QD ground state is beneficial \cite{Chang_2005,Benny_2011,Baier_2006}. 

\begin{acknowledgments}
The authors thank X. Cao for fruitful discussion and gratefully acknowledge the funding by the German Federal Ministry of Education and Research (BMBF) within the project Q.Link.X (16KIS0869) and QR.X (16KISQ015), the European Research Council (QD-NOMS GA715770) and the Deutsche Forschungsgemeinschaft (DFG, German Research Foundation) under Germany's Excellence Strategy (EXC- 2123) QuantumFrontiers (390837967).
\end{acknowledgments}

\bibliography{main_ref.bib}
\end{document}